\newcommand{\diracslash}[1]{#1\llap{/\kern2pt}}

\newcommand{\be}{\begin{equation}}
\newcommand{\ee}{\end{equation}}
\newcommand{\bea}{\begin{eqnarray}}
\newcommand{\eea}{\end{eqnarray}}
\newcommand{\ba}[1]{\begin{array}{#1}}
\newcommand{\ea}{\end{array}}

\newcommand{\bt}{\begin{tabular}}
\newcommand{\et}{\end{tabular}}

\newcommand{\beas}{\begin{eqnarray*}}
\newcommand{\eeas}{\end{eqnarray*}}

\documentclass[preprint,prd,aps,floats,nofootinbib,floatfix]{revtex4}

\DeclareSymbolFont{rsfs}{U}{rsfs}{m}{n}
\DeclareSymbolFontAlphabet{\mathrsfs}{rsfs}
\usepackage{amsmath}
\usepackage{slashed}
\usepackage{graphicx}
\usepackage{cleveref}
\usepackage{epstopdf}
\usepackage{xcolor}

\begin{document}

\title{In-medium masses and magnetic moments of decuplet baryons}

 \author{Harpreet Singh}
\email{harpreetmscdav@gmail.com}
\affiliation{Department of Physics, Dr. B R Ambedkar National Institute of Technology Jalandhar, 
 Jalandhar -- 144011, Punjab, India}
 \author{Arvind Kumar}
\email{iitd.arvind@gmail.com, kumara@nitj.ac.in}
\affiliation{Department of Physics, Dr. B R Ambedkar National Institute of Technology Jalandhar, 
 Jalandhar -- 144011, Punjab, India}
  \author{Harleen Dahiya}
\email{dahiyah@nitj.ac.in}
\affiliation{Department of Physics, Dr. B R Ambedkar National Institute of Technology Jalandhar, 
 Jalandhar -- 144011, Punjab, India}

\def\be{\begin{equation}}
\def\ee{\end{equation}}
\def\bearr{\begin{eqnarray}}
\def\eearr{\end{eqnarray}}
\def\zbf#1{{\bf {#1}}}
\def\bfm#1{\mbox{\boldmath $#1$}}
\def\hf{\frac{1}{2}}
\def\kp{\zbf k+\frac{\zbf q}{2}}
\def\km{-\zbf k+\frac{\zbf q}{2}}
\def\hwo{\hat\omega_1}
\def\hwt{\hat\omega_2}

\begin{abstract}

The magnetic moments of baryon decuplet are studied in 
vacuum as well as in the symmetric nuclear matter at finite temperature using  
a chiral SU(3) quark mean field model approach. The contributions coming from the valence quarks,  quark sea and the orbital angular momentum of the quark sea have been considered to calculate magnetic moment of decuplet baryons. The decuplet baryon masses decrease, whereas, the magnetic moments increase significantly with the rise of baryonic density of the symmetric nuclear medium. 
This is because of the reason that constituent quark magnetic moment and the quark spin polarizations show 
considerable variation in the nuclear medium especially in the low temperature and baryonic density regime. The increase is however quantitatively less as compared to the case of octet baryon members.
\end{abstract}

\maketitle

\section{Introduction}
The study of fundamental baryonic properties has been a major topic in particle physics from decades. The static properties of baryons such as magnetic moments, charge radii, form factors etc. have central importance for the study of internal structure of baryons. 
Magnetic moment is one of the important structural properties which gives better insight into baryon structure and can provide valuable understanding of mechanism of strong interaction at low energies, i.e., in the non perturbative regime of QCD (Quantum Chromo Dynamics) \cite{aliev1}. Since Coleman and Glashow \cite{hong1,hong12} predicted the magnetic moments of baryon octet, a lot of progress has been observed in both the theoretical paradigm and experimental verification for the study of baryon magnetic moments \cite{aman}. The magnetic moments of octet baryons in free space have been widely studied in different theoretical frameworks \cite{felix,wrb,hack,jg,jun,lks,ss}. On the experimental side seven of octet baryon magnetic moments are measured with around $1\%$ accuracy, whereas, the transition magnetic moment for $\Sigma^0\rightarrow\Lambda$ is known within $5\%$ precision \cite{contreras,contreras11}. However, for decuplet baryons the situation is quite different. Although the masses, decay aspects and other physical observables of some decuplet baryons are measured successfully, the magnetic moments of many decuplet baryons are yet to be determined \cite{milton,milton1}. The main reason behind this is the very short lifetime of most of decuplet baryons. The baryon $\Omega^-$ is an exception because of its substantially longer lifetime. There has recently been a renewed interest in experimental calculation of decuplet baryon magnetic moments. For example, the magnetic moment of $\Omega^-$ \cite{contreras12,contreras13} and $\Delta^{2+}$ has been measured experimentally \cite{contreras14}. Also, the magnetic moment of baryon $N^*(1535)$ will be studied at the experiments at Mainz Microtron (MAMI) \cite{aliev(2),aliev22,aliev23,aliev24} and Jefferson Lab \cite{aliev25}. An effort to measure the magnetic moment of $\Delta^+$ has also been realized at MAMI.  

On theoretical basis, the decuplet magnetic moments have been initially studied in non-relativistic simple quark model, which was further improved by including quark sea contribution \cite{aman12}, SU(3) symmetry breaking effects \cite{aman14,aman15,aman16} and orbital angular momentum of quarks \cite{aman13}. Later on the concept of effective mass to calculate magnetic moments of baryon magnetic moments was introduced in Ref. \cite{aman18}. In the recent studies, the decuplet magnetic moments have been predicted in models like relativistic quark model \cite{aman20,aman21}, QCD sum rule \cite{aman27,aman28,aman29}, light cone QCD sum rule \cite{aman26}, Skyrme model \cite{aman30,aman31}, effective mass scheme \cite{aman24,aman25}, chiral perturbation theory \cite{aman35,aman36}, chiral quark soliton model \cite{aman32,aman33,aman34}, lattice QCD \cite{aman37,aman38,aman39,aman40} etc. 

The study of medium modification of the baryonic properties is gaining interest day by day because of its importance for astrophysical studies and heavy ion collision studies. The experimental facilities such as FAIR at GSI Germany \cite{friese,gsi}, Cooling Storage Ring (CSR) at HIRFL in China \cite{zhan}, Radioactive Ion Beam (RIB) Factory at RIKEN in Japan \cite{yano}, SPIRAL2/GANIL in France \cite{ganil} and Facility for Rare Isotope Beams (FRIB) in United States \cite{frib}  etc. have planned different measurements with main focus at medium modification of baryon properties. For experimental realization of medium modification of decuplet magnetic moments, one requires a consistent  theoretical model which can provide resourceful predictions of magnetic moments in the baryonic medium at finite temperature.
Recently, medium modification of decuplet baryon properties as a function of temperature were discussed in \cite{azizi}. However, the dependence of magnetic moments of baryon decuplet members on baryonic density at finite  temperature of nuclear medium has not been discussed in the available literature. The medium modification of octet baryon magnetic moments was studied in Refs. \cite{ryu1,ryu2} using quark meson coupling model and modified quark meson coupling model. We have used the chiral SU(3) quark mean field model to calculate magnetic moments of baryon octet members in symmetric nuclear matter and asymmetric nuclear matter at finite temperature \cite{happy,happy1}. 
The thermodynamic consistency of mean field models make them reliable for calculation of in-medium baryonic properties \cite{serot2}. In the light of above developments, the study of magnetic moments of decuplet baryons in mean field approximation method  by including quark degrees of freedom in hot and dense nuclear medium will be quite interesting and is the main goal of the present work.

The study of medium modification of decuplet magnetic moments can be an important step forward in the exploration of behavior of structural properties of resonance particles in the presence of the nuclear medium. 
In \cref{masscalcub}, we briefly outline the model used to calculate in-medium decuplet baryon magnetic moments. 
In section \ref{results}, we present the numerical calculations and results. The summary of present study is given in section \ref{summ}. 
     

\section{Decuplet Baryons in Symmetric Nuclear Matter}
\label{masscalcub} 
In order to investigate medium modified magnetic moments of decuplet baryons we use the idea of chiral quark model \cite{sw,am}. The chiral quark model not only include the effects of quark confinement and chiral symmetry breaking but allows the use of effective quark masses in place of constituent quark masses. The chiral quark model advocates the existence of Goldstone bosons (GBs), generated from the valence quarks. The GB thus generated is further splits into quark-antiquark pair and leads to generation of quark sea \cite{chengsu3,cheng1,song}. The constituent quark wave function of chiral quark model gets perturbed because the valence quarks undergo chromodynamic spin-spin forces. Hence, the total magnetic moment of decuplet baryon consists of contributions from valence quarks and quark sea and hence given as \cite{cheng,har}
 \be
 \mu_B^{*}= \mu_B^{\text{val}}+\mu_B^{\text{sea}^{'}}, \label{magtotal}
 \ee
where $\mu_B^{\text{val}}$ is the contribution from valence quark given as   
 \be
\mu_B^{{\rm val}}=\sum_{q=u,d,s} {\Delta q^{{\rm val}}\mu_q^{*}}. ~~~~
\ee
In above, $\Delta q^{{\rm val}}$ is the valence quark polarization which can be calculated as done in \cite{har} and $\mu_q^*$ represents in-medium magnetic moment of constituent quark.
Further, in \cref{magtotal}, $\mu_B^{\text{sea}^{'}}$ represents the contribution from quark sea, which comprises of contributions from magnetic moment $\mu_B^{\text{sea}}$ of a quark in the quark sea as well as the magnetic moment $\mu_B^{\text{orbital}}$ resulting from orbital angular momentum of the quark in the quark sea and is written as
\be
\mu_B^{\text{sea}^{'}}=\mu_B^{\text{sea}}+\mu_B^{\text{orbital}}.
\ee
The contribution from quarks in the quark sea is written as 
\be
\mu_B^{{\rm sea}}=\sum_{q=u,d,s} {\Delta q^{{\rm sea}}\mu_q^*}. ~~~     \label{mag}
\ee
The sea quark polarization $\Delta q^{{\rm sea}}$ is calculated in terms of symmetry breaking parameters $\varepsilon$, $\varpi$ and $\tau$. For example, in case of $\Delta^{2+}$ the quark sea polarizations for different quark flavors are given as \cite{har,nmc}
\begin{eqnarray}
&&\Delta u^{\rm sea}=-a\left(6+3\varepsilon^2+\varpi^2+2\tau^2\right),  \nonumber \\
&&\Delta d^{\rm sea}=-3a,     \nonumber \\
&&\Delta s^{\rm sea}=-3a\varepsilon^2.
\end{eqnarray}
The details of symmetry breaking parameters and the constant `$a$' can be found in our recent work in \cite{happy}, where we calculated these parameters for octet baryons. The symmetry breaking parameters are found to get modify in the nuclear matter, leading to medium modification of quark sea polarizations. The contribution from orbital angular momentum of quark sea is written as 
\begin{equation}
\mu_{B}^{\rm orbital}=\Delta u^{\rm val}[\mu\left(u^{+}\rightarrow\right)]+\Delta d^{\rm val}[\mu\left(d^{+}\rightarrow\right)]+\Delta s^{\rm val}[\mu\left(s^{+}\rightarrow\right)].
\end{equation} 
The transition magnetic moments ($\mu(q^{+}\rightarrow)$) are essentially the same as in \cite{happy}. These transition magnetic moments also get modified in the nuclear medium. 

The mass adjusted magnetic moments of constituent quarks appering in \cref{magtotal} are given as \cite{aarti,har2,gupta} 
\begin{equation}
 \mu_d^* =-\left(1-\frac{\Delta M}{M_B^*}\right),~~  \mu_s^*=-\frac{m_u^{*}}{m_s^{*}}\left(1-\frac{\Delta M}{M_B^*}\right),~~  \mu_u^*=-2\mu_d^* ,\label{magandmass}         
\end{equation}     
where $\Delta M=M_{B}^{*}-M_{\rm vac}$ is the mass difference between the medium modified mass of baryon $M_B^*$ and the experimental vacuum mass of baryon ($M_{\rm vac}$). Further, $m_u^{*}$ and $m_s^{*}$ represent the medium modified mass of $u$ and $s$ quarks, respectively. We obtain the medium modified masses of decuplet baryons and constituent quarks using chiral SU(3) quark mean field model approach. In the chiral SU(3) quark mean field model, the constituent quark masses and energies are obtained using the values of scalar meson mean fields $\sigma$ and $\zeta$, which are obtained by solving coupled equations of motion for these mesons in the nuclear medium in a self-consistent manner \cite{wang,happy,wang2,wang3,wang4}. The medium modified mass of baryon is written as   
\begin{equation}
M_B^*=\sqrt{E_B^{*2}- <p_{B \, \text{cm}}^{*2}>}\,, \label{baryonmass}
\end{equation} 
where $E_B^{*}$ represents the in-medium energy of $B^{th}$ decuplet baryon given as
\begin{equation} 
E_B^*=\sum_{q=u,d,s} n_{qB}e_q^*+E_{B \, \text{spin}},    \label{energy}
\end{equation}
where $n_{qB}$ represents the number of quarks of type $q$ in $B^{th}$ decuplet baryon. The $e_q^*$ is the in-medium energy of constituent quark. Further, $E_{B \, \text{spin}}$ is the correction to baryon energy coming from spin-spin interaction between constituent quarks and is determined by fitting the vacuum mass of particular decuplet baryon. The values used in the present work are:
\begin{eqnarray}
&&E_{\Delta \, \text{spin}} = - 214 \,\,\, {\rm MeV}\,,
~E_{\Omega^- \, \text{spin}} = - 192 \,\,\, {\rm MeV}\,,
~E_{\Sigma^* \, \text{spin}} =  - 201 \,\, {\rm MeV}\,, \nonumber \\
~~&&E_{\Xi^{*0} \, \text{spin}} =  - 193 \,\, {\rm MeV}\,,
~E_{\Xi^{*-} \, \text{spin}} =  - 190 \,\, {\rm MeV}\, .
\end{eqnarray}
It is important to note that $E_{B \, \text{spin}}$ causes the mass difference between octet and decuplet baryons with similar quark content and it is found to remain invariant with respect to changes in baryonic density and temperature of the medium. In \cref{baryonmass}, $<p_{i \, \text{cm}}^{*2}>$ is the spurious center of mass motion \cite{happy,barik1,barik2}.

\section{Numerical results} \label{results}
\begin{figure}
\resizebox{0.7\textwidth}{!}{
\includegraphics{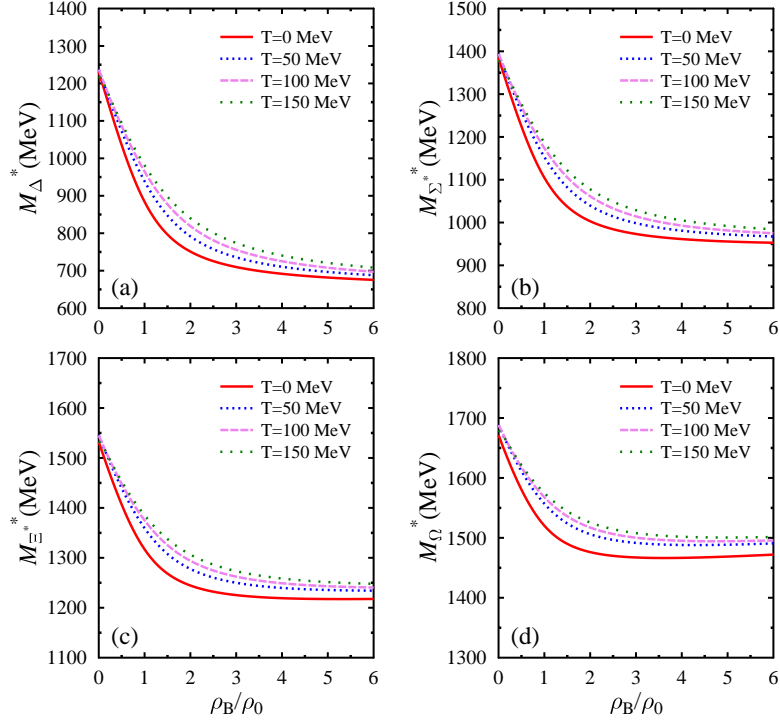} }
\caption{{Effective masses of decuplet baryons (at T=0, 50, 100 and 150 MeV) versus baryonic density (in units of nuclear saturation density $\rho_0$).} } \label{bmasst}
\end{figure}

In this section we present the results of our calculations of masses and magnetic moments of decuplet baryons at different  baryonic densities and/or temperature of the symmetric nuclear medium. In \cref{bmasst}, we have plotted the medium modified decuplet baryon masses ($M_B^*$, $B=\Delta, \Sigma^*, \Xi^* , \Omega$) versus baryonic density of the nuclear medium at different temperatures ($0, 50, 100$ and $150$ MeV). We find that at a given temperature, with the increase in baryonic density of the medium, the decuplet baryon members show a decrease in their effective masses. The non-strange baryons show very large decrease in their masses for the baryonic densities ranging from $\rho_B=0$ to $\rho_B=2\rho_0$. However, for baryonic densities more than $2\rho_0$, the decrease in effective masses becomes slow. On the other hand, in the case of strange baryons at given temperature, the decrease in effective baryon masses with the increase of density is small as compared to non-strange baryons especially in low density region. 



\begin{figure}
\resizebox{0.6\textwidth}{!}{
\includegraphics{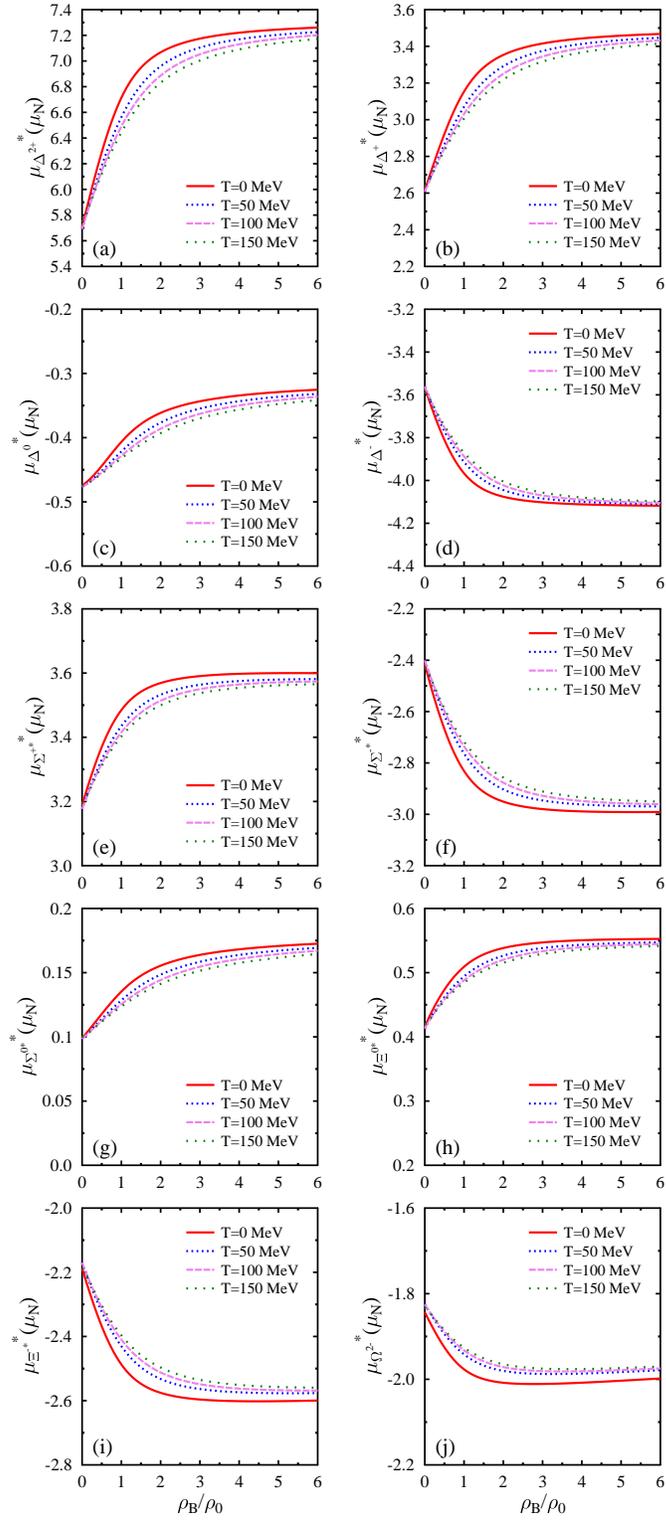} }
\caption{{Magnetic moment of decuplet baryons (at T = 0, 50, 100 and 150 MeV) versus baryonic density (in units of nuclear saturation density $\rho_0$).} } \label{mmt}
\end{figure}
\begin{figure}
\resizebox{0.6\textwidth}{!}{
\includegraphics{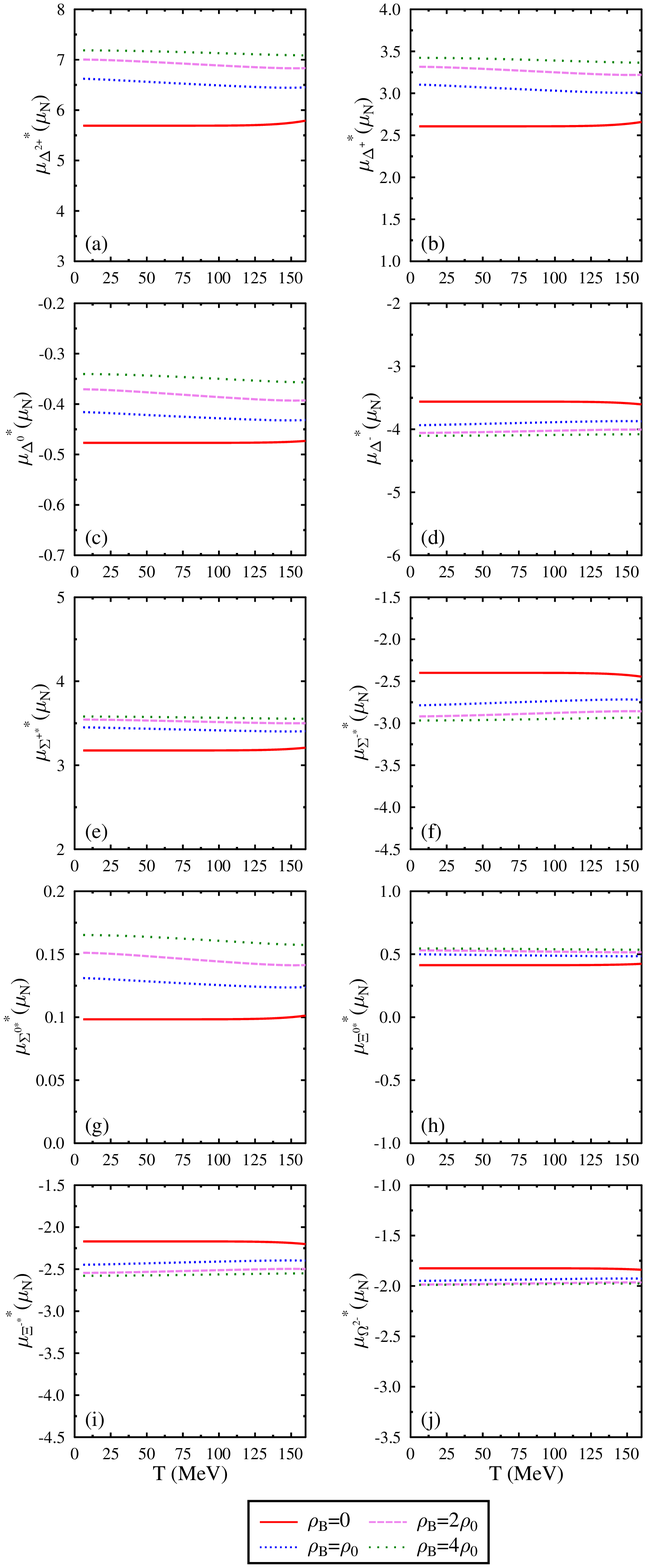} }
\caption{{Magnetic moments of baryons as a function of temperature at $\rho_B=0,\rho_0,2\rho_0$ and $4\rho_0$.} } \label{mmpt}
\end{figure}

  
For example, the $\Delta$ baryons show a decrease of $29\%$ in their effective masses as compared to their vacuum masses  with the increase of baryonic density from $\rho_B=0$ to $\rho_B=\rho_0$ at zero temperature. 
In case of $\Sigma^*$, $\Xi^*$ and $\Omega$ baryons there is a decrease of $21.3\%$, $15.1\%$ and $10\%$ in the respective effective masses as compared to their vacuum masses. 
Further, at zero temperature, for the increase of density from $\rho_0$ to $2\rho_0$, the percentage decrease in baryon masses are found to be $22\%$, $11\%$, $7\%$ and $4.7\%$ for $\Delta$, $\Sigma^*$, $\Xi^*$ and $\Omega$ baryons, respectively. In \cref{bmasst} one can see that for given temperature of the medium, 
for densities more than $2\rho_0$, $M_{\Delta}^*$ shows a small decrease and $M_{\Sigma^*}^*$ and $M_{\Xi^*}^*$ show very small decrease, $M_{\Omega}^*$ becomes almost constant and shows very small increase for densities more than $4\rho_0$. This can be understood in terms of the fact that effective masses of decuplet baryons depend on effective masses of their  constituent quarks which in turn depend on scalar meson fields $\sigma$ and $\zeta$. The behavior of meson fields in the chiral SU(3) quark mean field model, with the increase of baryonic density and for given temperature of medium has been thoroughly discussed in \cite{happy}. It was observed that the $\sigma$ meson field couple strongly to the $u$ and $d$ quarks and $\zeta$ meson field couple to the strange quark. Further, at given temperature, the magnitude of $\sigma$ meson field shows a rapid decrease with the increase of baryonic density of the medium, whereas, the magnitude of $\zeta$ meson field shows a very slow decrease with the increase of baryonic density. For densities more than $4\rho_0$, the magnitude of $\zeta$ field shows an increase, which is due to deconfinement phase transition at these densities \cite{abu}. As strange quark's effective mass depends only on $\zeta$ meson field because of absence of coupling between strange quarks and $\sigma$ field, hence, the effective mass of strange quark decreases upto $4\rho_0$ and then shows a small increase. Therefore, the effective mass of $\Omega$ baryon shows an increase for baryonic density more than $4\rho_0$ at given temperature of the nuclear medium.           

Comparing the increase and/or decrease of medium modified masses of decuplet baryons and octet baryons (calculated in \cite{happy}) as a function of baryonic density, at given temperature of the medium, we find that decuplet baryons with same quark content as that of octet baryon members show a small decrease in their effective masses as compared to the case of their corresponding octet baryon members. For example, at zero temperature, with the increase of baryonic density from zero to the nuclear saturation density, the nucleons show a decrease of $41\%$ in their effective masses as compared to their vacuum mass, whereas, $\Delta$ baryons with same quark content show a decrease of $29\%$. This is due to increased spin energy `$E_{\rm spin}$' contribution to the effective baryon mass in case of decuplet baryons leading to less contribution from increasing and/or decreasing effective masses of constituent quarks to the effective masses of decuplet baryons.


One can also observe that, at given baryonic density, the effective masses of baryons show an increase as a function  of temperature of the medium. For example, at $\rho_B=0$, effective mass of $\Delta$, $\Sigma^*$, $\Xi^*$ and $\Omega$ baryons increase by $5$ MeV, $9$ MeV, $13$ MeV and $16$ MeV, respectively, for the rise of temperature from 0 to 100 MeV. This shows that at zero baryonic density, with the rise of strangeness content of the decuplet baryon, the effective mass of baryon shows an increase as a function of temperature of the medium.

  
On the other hand, at nuclear saturation density, with the increase of temperature from  0 to 100 MeV, the effective masses of $\Delta$, $\Sigma^*$, $\Xi^*$ and $\Omega$ baryons increase by $80.7$ MeV, $69$ MeV, $58$ MeV and $43$ MeV, respectively. This shows that the effect of rise of temperature on effective mass of non-strange decuplet baryons is more prominent at finite baryonic density as compared to the case of $\rho_B=0$, whereas, at finite baryonic density, the strange baryons shows less increase in their effective masses as a function of temperature of the medium as compared to non-strange baryons.     

Henceforth we discuss the results for in-medium magnetic moments of decuplet baryons. In \cref{mmt}, we have plotted  magnetic moment of decuplet baryons as a function of baryonic density at different temperatures ($0, 50, 100$ and $150$ MeV). In \cref{mmval,mmvaldifft}, we have tabulated the results for in-medium magnetic moment of decuplet baryons, at zero and finite (100 MeV) temperature, respectively. 
To compare the vacuum magnetic moments of baryons calculated in the present work with the available literature we have given \cref{mmval1}. As can be seen from this table, the results for vacuum magnetic moments of decuplet baryons in the present work are comparable to those obtained in other theoretical approaches, especially to those obtained through Large-$N_c$ approach in \cite{aarti24} and $\chi{\rm CQM}$ model approach used in \cite{aarti}.  The present results are very close to those obtained in Ref. \cite{aarti}, however, with the exceptions of $\mu_{\Sigma^{*-}}^*$ and $\mu_{\Sigma^{*0}}^{*}$. The $\chi{\rm CQM}$ using constituent quark masses has been used to calculate vacuum magnetic moments of decuplet baryons in Ref. \cite{aarti} and the vacuum masses of constituent quarks have been taken as $m_u=m_d=330$ MeV and $m_s=500$ MeV. In the present work we have used  $m_u=m_d=313$ MeV and $m_s=490$ MeV as well as the symmetry breaking parameters have been calculated through their explicit dependence upon baryon masses.

From \cref{mmval,mmvaldifft} one can see that at zero as well as finite temperature, with the rise of baryonic density, the contributions from valence quarks and quark sea to the total effective magnetic moments of baryons increase. The  exceptions are in case of $\Delta^-$ and $\Sigma^{*-}$ where the contribution from quark sea shows a decrease in magnitude with rise of density from $\rho_B=0$ to $\rho_0$ and then shows an increase from $\rho_0$ to $4\rho_0$. The contribution coming from the orbital angular momentum part of quark sea decreases with the rise of baryonic density at zero as well as finite temperature, with exceptions in case of $\mu_{\Delta^-}^*$, $\mu_{\Sigma^{*-}}^*$ and $\mu_{\Sigma^{*0}}^{*}$. In case of $\Delta^-$ and $\Sigma^{*-}$, $\mu^{\rm orbital}$ changes of sign at $\rho_B=4\rho_0$ as compared to its value at $\rho_B=\rho_0$, whereas, in case of $\Sigma^{*0}$ the $\mu^{\rm orbital}$ show an increase in magnitude with the increase of baryonic density of nuclear medium. The exceptional behavior of $\mu^{\rm sea}$ and $\mu^{\rm orbital}$ in case of $\Sigma^{*0}$, $\Delta^-$ and $\Sigma^{*-}$, is due to exceptional behavior of spin wave functions of these baryons in the presence of medium.

Further, we observe that with the rise of strangeness content of decuplet baryon, the amount of variations in $\mu^{\rm sea}$ and $\mu^{\rm orbital}$ decrease. At given temperature of medium, the total effective magnetic moments of strange baryons also show less variation as a function of density, as compared to non-strange baryons. The reason behind the observed behavior of total effective magnetic moments of baryons is their direct dependence on effective masses of respective baryons. As discussed earlier, the effective mass of non-strange baryons show large increase with the increase of baryonic density of the nuclear medium, at given temperature, whereas, in case of strange baryons this increase in effective mass is comparatively less. Hence, $\mu_{u}^*$ and $\mu_{d}^*$ decrease more rapidly with the rise of density of medium as compared to $\mu_{s}^*$, and thus, the effective magnetic moments of the baryons with larger `$s$' quark content show less decrease as a function of density.       
 
On comparing the values in tables \ref{mmval} and \ref{mmvaldifft}, we observe that the temperature negligibly effects the  effective magnetic moment of decuplet baryons in vacuum ($\rho_B=0$).
This is because of very small effect of temperature on $\mu^{\rm val}$, $\mu^{\rm sea}$ and $\mu^{\rm orbital}$. On the other hand, in the baryonic medium (finite densities), there is a countable decrease in $\mu^{\rm sea}$ along with $\mu^{\rm val}$, whereas $\mu^{\rm orbital}$ increases. This leads to decrease in the values of total effective magnetic moments of baryons with the rise of temperature from T = 0 to T = 100 MeV. For example, in case of $\Delta^{2+}$ at $\rho_B=\rho_0$, for rise of temperature from T = 0 to 100 MeV, $\mu^{\rm val}$ and $\mu^{\rm sea}$ decrease by $0.393 \mu_N$, $0.070 \mu_N$ respectively, whereas, $\mu^{\rm orbital}$ increases by $0.099 \mu_N$. Due to this $\mu_{\Delta^{2+}}^*$ decreases from $6.713 \mu_N$ at T = 0 MeV to $6.491 \mu_N$ at T = 100 MeV. At $\rho_B=4\rho_0$ the situation is somewhat different, as the amount of decrease in $\mu^{\rm val}$ is small as compared to the case of $\rho_0$. 

 
 \begin{table*}[ht]
 \resizebox{.95\hsize}{!}{ 
 \begin{tabular}{|c|c|c|c|c|c|c|c|c|c|c|c|c|c|}  
 \hline 
 & Data \cite{experiment,aman16} & \multicolumn{4}{c|}{$\rho_B=0$} & \multicolumn{4}{c|}{$\rho_B=\rho_0$} & \multicolumn{4}{c|}{$\rho_B=4\rho_0$}   \\ 
 \hline
 & $\mu_B$ (In free space) & $\mu^{\text{val}}$ & $\mu^{\text{sea}}$ & $\mu^{\text{orbital}}$ & $\mu_B^{*}$ & $\mu^{\text{val}}$ & $\mu^{\text{sea}}$ & $\mu^{\text{orbital}}$ & $\mu_B^{*}$ & $\mu^{\text{val}}$ & $\mu^{\text{sea}}$ & $\mu^{\text{orbital}}$ & $\mu_B^{*}$  \\ 
 \hline 
 $\mu_{\Delta^{2+}}^*(\mu_N)$ & $4.5-7.5$ & $6$ & $-1.068$ & $0.771$ & $5.703$ & $7.695$ & $-1.329$ & $0.348$ & $6.713$ & $8.631$ & $-1.514$ & $0.103$ & $7.220$   \\ 
 \hline 
 $\mu_{\Delta^+}^*(\mu_N)$ & $--$ & $3$ & $-0.675$ & $0.289$ & $2.614$ & $3.847$ & $-0.857$ & $0.163$ & $3.153$ & $4.316$ & $-0.969$ & $0.096$ & $3.443$   \\ 
 \hline 
 $\mu_{\Delta^0}^*(\mu_N)$ & $--$ & $0$ & $-0.282$ & $-0.192$ & $-0.474$ & $0$ & $-0.384$ & $-0.022$ & $-0.407$ & $0$ & $-0.424$ & $0.090$ & $-0.334$  \\ 
 \hline 
 $\mu_{\Delta^-}^*(\mu_N)$ & $--$ & $-3$ & $0.110$ & $-0.674$ & $-3.563$ & $-3.847$ & $0.088$ & $-0.207$ & $-3.967$ & $-4.316$ & $0.120$ & $0.084$ & $-4.112$   \\ 
 \hline 
 $\mu_{\Sigma^{*+}}^*(\mu_N)$ & $--$ & $3.361$ & $-0.681$ & $0.507$ & $3.187$ & $4.041$ & $-0.786$ & $0.229$ & $3.485$ & $4.389$ & $-0.859$ & $0.067$ & $3.598$   \\ 
\hline 
 $\mu_{\Sigma^{*-}}^*(\mu_N)$ & $--$ & $-2.221$ & $0.104$ & $-0.297$ & $-2.414$ & $-2.84$ & $0.1$ & $-0.091$ & $-2.831$ & $-3.156$ & $0.13$ & $0.037$ & $-2.989$   \\  
\hline 
 $\mu_{\Sigma^{*0}}^*(\mu_N)$ & $--$ & $0.361$ & $-0.288$ & $0.026$ & $0.098$ & $0.434$ & $-0.343$ & $0.044$ & $0.135$ & $0.472$ & $-0.364$ & $0.061$ & $0.168$   \\  
 \hline 
 $\mu_{\Xi^{*0}}^*(\mu_N)$ & $--$ & $0.722$ & $-0.294$ & $-0.013$ & $0.415$ & $0.824$ & $-0.308$ & $-0.006$ & $0.509$ & $0.870$ & $-0.317$ & $-0.002$ & $0.550$   \\ 
 \hline 
 $\mu_{\Xi^{*-}}^*(\mu_N)$ & $--$ & $-2.272$ & $0.098$ & $-0.013$ & $-2.187$ & $-2.592$ & $0.111$ & $-0.006$ & $-2.487$ & $-2.738$ & $0.139$ & $-0.002$ & $-2.602$   \\ 
 \hline 
 $\mu_{\Omega}^*(\mu_N)$ & $-2.02\pm 0.005$ & $-1.917$ & $0.093$ & $-0.019$ & $-1.843$ & $-2.090$ & $0.122$ & $-0.009$ & $-1.977$ & $-2.152$ & $0.147$ & $-0.004$ & $-2.008$   \\ 
 \hline 
 \end{tabular}}
\caption{Effective magnetic moments of decuplet baryons at T = 0 MeV and $\rho_B=0, \rho_0$ and $4\rho_0$.} \label{mmval}
\end{table*} 

 \begin{table*}[hb]
 \resizebox{.95\hsize}{!}{
  \begin{tabular}{|c|c|c|c|c|c|c|c|c|c|c|c|c|}  
 \hline 
 &  \multicolumn{4}{c|}{$\rho_B=0$} & \multicolumn{4}{c|}{$\rho_B=\rho_0$} & \multicolumn{4}{c|}{$\rho_B=4\rho_0$}   \\ 
 \hline
 & $\mu^{\text{val}}$ & $\mu^{\text{sea}}$ & $\mu^{\text{orbital}}$ & $\mu_B^{*}$ & $\mu^{\text{val}}$ & $\mu^{\text{sea}}$ & $\mu^{\text{orbital}}$ & $\mu_B^{*}$ & $\mu^{\text{val}}$ & $\mu^{\text{sea}}$ & $\mu^{\text{orbital}}$ & $\mu_B^{*}$  \\ 
 \hline 
 $\mu_{\Delta^{2+}}^*(\mu_N)$ & $5.975$ & $-1.06$ & $0.777$ & $5.692$ & $7.302$ & $-1.259$ & $0.448$ & $6.491$ & $8.468$ & $-1.483$ & $0.145$ & $7.130$   \\ 
 \hline 
 $\mu_{\Delta^+}^*(\mu_N)$ & $2.988$ & $-0.671$ & $0.291$ & $2.608$ & $3.651$ & $-0.812$ & $0.192$ & $3.031$ & $4.234$ & $-0.950$ & $0.106$ & $3.390$   \\ 
\hline 
 $\mu_{\Delta^0}^*(\mu_N)$ & $0$ & $-0.283$ & $-0.194$ & $-0.476$ & $0$ & $-0.365$ & $-0.063$ & $-0.428$ & $0$ & $-0.417$ & $0.067$ & $-0.349$   \\ 
\hline 
 $\mu_{\Delta^-}^*(\mu_N)$ & $-2.988$ & $0.105$ & $-0.679$ & $-3.562$ & $-3.651$ & $0.082$ & $-0.318$ & $-3.888$ & $-4.234$ & $0.116$ & $0.028$ & $-4.090$   \\  
 \hline 
 $\mu_{\Sigma^{*+}}^*(\mu_N)$ & $3.339$ & $-0.674$ & $0.512$ & $3.177$ & $3.874$ & $-0.754$ & $0.295$ & $3.415$ & $4.313$ & $-0.843$ & $0.095$ & $3.564$  \\ 
 \hline 
 $\mu_{\Sigma^{*-}}^*(\mu_N)$ & $-2.202$ & $0.101$ & $-0.299$ & $-2.401$ & $-2.688$ & $0.093$ & $-0.140$ & $-2.736$ & $-3.087$ & $0.1256$ & $0.013$ & $-2.948$   \\ 
 \hline 
 $\mu_{\Sigma^{*0}}^*(\mu_N)$ & $0.358$ & $-0.287$ & $0.026$ & $0.098$ & $0.416$ & $-0.330$ & $0.039$ & $0.125$ & $0.463$ & $-0.359$ & $0.056$ & $0.161$   \\ 
 \hline 
 $\mu_{\Xi^{*0}}^*(\mu_N)$ & $0.716$ & $-0.290$ & $-0.012$ & $0.413$ & $0.796$ & $-0.301$ & $-0.006$ & $0.488$ & $0.856$ & $-0.312$ & $-0.003$ & $0.539$   \\ 
 \hline 
 $\mu_{\Xi^{*-}}^*(\mu_N)$ & $-2.253$ & $0.096$ & $-0.012$ & $-2.170$ & $-2.506$ & $0.103$ & $-0.006$ & $-2.410$ & $-2.693$ & $0.134$ & $-0.003$ & $-2.562$   \\ 
 \hline 
 $\mu_{\Omega}^*(\mu_N)$ & $-1.898$ & $0.091$ & $-0.019$ & $-1.825$ & $-2.036$ & $0.113$ & $-0.010$ & $-1.932$ & $-2.119$ & $0.143$ & $0.005$ & $-1.981$   \\ 
 \hline 
 \end{tabular}}
\caption{Effective magnetic moments of decuplet baryons at T = 100 MeV and $\rho_B=0, \rho_0$ and $4\rho_0$.} \label{mmvaldifft}
\end{table*}

\begin{table*}[ht]
 \resizebox{.95\hsize}{!}{ 
 \begin{tabular}{|c|c|c|c|c|c|c|c|c|c|}  
 \hline
 & Present Work & $\chi {\rm QM}$ \cite{aman16} & $\chi {\rm PT}$\cite{aarti20} & CQSM \cite{aarti18} & LQCD \cite{aarti22} & CBM \cite{aarti23} & Large ${\rm N_c}$ \cite{aarti24} & $\chi {\rm QMEC}$ \cite{aarti26} & $\chi {\rm CQM}$ with $\mu_{q}^{eff}$   \cite{aarti} \\ 
 \hline 
 $\mu_{\Delta^{2+}}(\mu_N)$ & $5.703$ & $5.300$ & $5.390$ & $4.730$ & $5.240$ & $4.520$ & $5.900$ & $6.930$ & $5.820$    \\ 
 \hline 
 $\mu_{\Delta^+}(\mu_N)$ & $2.614$ & $2.580$ & $2.383$ & $2.190$ & $0.95$ & $2.120$ & $2.900$ & $3.470$ & $2.630$    \\ 
 \hline 
 $\mu_{\Delta^0}(\mu_N)$ & $-0.474$ & $-0.130$ & $-0.625$ & $-0.350$ & $-0.035$ & $-0.290$ & $--$ & $0$ & $-0.550$   \\ 
 \hline 
 $\mu_{\Delta^-}(\mu_N)$ & $-3.563$ & $-2.850$ & $-3.632$ & $-2.900$ & $-2.980$ & $-2.690$ & $-2.900$ & $-3.470$ & $-3.750$    \\ 
 \hline 
 $\mu_{\Sigma^{*+}}(\mu_N)$ & $3.187$ & $2.880$ & $2.519$ & $2.520$ & $1.270$ & $2.630$ & $3.300$ & $4.120$ & $3.090$    \\ 
\hline 
 $\mu_{\Sigma^{*-}}(\mu_N)$ & $-2.414$ & $-2.550$ & $-3.126$ & $-2.690$ & $-1.880$ & $-2.48$ & $-2.800$ & $-3.060$ & $-3.070$    \\  
\hline 
 $\mu_{\Sigma^{*0}}(\mu_N)$ & $0.098$ & $0.170$ & $-0.303$ & $-0.080$ & $0.330$ & $0.080$ & $0.300$ & $0.530$ & $0.018$    \\  
 \hline 
 $\mu_{\Xi^{*0}}(\mu_N)$ & $0.415$ & $0.470$ & $0.149$ & $0.190$ & $0.160$ & $0.440$ & $0.650$ & $1.100$ & $0.460$    \\ 
 \hline 
 $\mu_{\Xi^{*-}}(\mu_N)$ & $-2.187$ & $-2.250$ & $-2.596$ & $-2.480$ & $-0.620$ & $-2.270$ & $-2.300$ & $-2.610$ & $-2.550$    \\ 
 \hline 
 $\mu_{\Omega}(\mu_N)$ & $-1.843$ & $-1.950$ & $-2.042$ & $-2.270$ & $--$ & $-2.060$ & $-1.940$ & $-2.130$ & $-2.090$    \\ 
 \hline 
 \end{tabular}}
\caption{Vacuum magnetic moments of decuplet baryons in different models.} \label{mmval1}
\end{table*}
       
In order to explore the effect of temperature on the magnetic moments more thoroughly, we have plotted the effective magnetic moments of baryons as a function of temperature at $\rho_B=0$, $\rho_0, 2\rho_0$ and $4\rho_0$ in  \cref{mmpt}. In the high baryonic density range, the effective values of magnetic moment of baryons show a negligible change as a function of temperature as compared to that in the lower baryonic density range. 
This is due to  \cref{mag,magandmass} which show that the effective magnetic moment of baryons are inversely proportional to the medium modified values of respective baryon masses. At $\rho_B=0$, the effective baryon masses remain almost same with the rise of temperature upto critical temperature as they are affected by the variation of thermal distribution functions of nucleons of the medium on the self energy of constituent quarks only, leading to decrease in the effective baryon masses, and hence, increasing  the effective magnetic moment of baryons. However, in the baryonic medium (finite densities), the higher momentum states also starts contributing in opposite sense, and hence, the effective magnetic moments start decreasing \cite{AMU}. Further, in the higher density regime ($4\rho_0$ or more) and higher temperature, because of the second order phase transition, the effective magnetic moment of baryons become insensitive to the variation in effective mass of baryons. This observation is further justified by those expected for octet baryons in Ref. \cite{happy} using the same chiral SU(3) quark mean field model and \cite{ryu1} 
using modified quark meson coupling model. 

\section{Summary} \label{summ}
To summarize,  we have investigated the effect of baryonic density and temperature of the nuclear medium on the decuplet baryon masses as well as magnetic moments using the chiral SU(3) quark mean field model. The non-strange baryons show a significant decrease with the rise of baryonic density of the nuclear medium especially in the low density regime. At $\rho_B=0$, with the rise of temperature, the effective masses of non-strange decuplet baryons show a small increase from their vacuum values as compared to strange baryons. However, at finite densities the non-strange baryons show comparatively large increase in effective masses as compared to strange baryons.  

Considering the baryonic magnetic moments to consist of contributions from valence quarks, quark sea and orbital angular momentum of the quark sea, we find that contribution from valence quarks and orbital angular momentum of quark sea are of the same sign, whereas the quark sea effect provides contribution of opposite sign. It is interesting to observe that the orbital angular momentum of quark sea contributes significantly to the magnetic moment particularly at low baryonic densities of the medium. However, for a given temperature and at higher densities, the opposite contribution from orbital angular momentum part of quark sea becomes very small.     

The rise of temperature of the nuclear medium decreases the effect of baryonic density on the magnetic moment of baryons. The   strange baryons show very slow variation in their magnetic moments as a function of baryonic density and/or temperature as compared to those of baryons with zero strangeness content. This is due to the dependence of baryonic magnetic moments on  in-medium strange quark mass, which shows very slow variation in symmetric nuclear matter. Further, for baryonic densities $4\rho_0$ or more the magnetic moments of baryons hardly show any variation as a function of temperature indicating second order phase transition at higher densities \cite{wang neu}.

The results for vacuum as well as in-medium magnetic moments of decuplets can be further improved by considering the  effects from relativistic exchange currents \cite{mg}, pion cloud contributions \cite{thomas} and the effects of confinement \cite{har2} etc. 
The present results can be very useful for the predictions of decuplet magnetic moments not only in free space but also in the presence of hadronic media created in heavy ion collision experiments focused at structural study of baryons, such as FAIR \cite{gsi}, RHIC \cite{rhic} and NICA \cite{nica}. Also, the present results can be of relevance for the Baryon Antibaryon Symmetry Experiment (BASE) \cite{base} at LHC.


\begin{thebibliography}{1}
\section*{References}
\bibitem{aliev1}
T. M. Aliev, V. S. Zamiralov, Advances in High Energy Physics {\bf 2015}, 406875 (2015).  

\bibitem{hong1}
S. T. Hong, Phys. Rev. D {\bf 76}, 094029 (2007).

\bibitem{hong12}
S. Coleman, S. L. Glashow, Phys. Rev. Lett. {\bf 6}, 423 (1961).

\bibitem{aman}
A. Kaur, A. Upadhyay Eur. Phys. J. A. {\bf 52}, 105 (2016).

\bibitem{felix}
F. Schlumpf, Phys. Rev. D {\bf 48} , 4478 (1993).

\bibitem{wrb}
W. R. B. de Araujo \textit{et al.}, Brazilian Journal of Physics {\bf 34} , 871  (2004).

\bibitem{hack}
E. J. Hackett-Jones, D. B. Leinweber, A. W. Thomas, Phys. Lett. B {\bf 489}, 143 (2000).

\bibitem{jg}
J. G. Contreras, R. Huerta, Revista Mxicana De Fisica {\bf 50} , 490 (2004).

\bibitem{jun}
H. E. Jun, Dong Yu-Bing, Commun. Theor. Phys. {\bf 43} , 139 (2005).

\bibitem{lks}
L. K. Sharma, C. Mai, J. Sci {\bf 34} , 13 (2007).

\bibitem{ss}
S. Sahu, Revista Mxicana De Fisica {\bf 48}, 48 (2002).

\bibitem{contreras}
J. G. Contreras, R. Huerta, L. R. Quintero, REVISTA MEXICANA DE FI´SICA {\bf 50}, 490 (2004).


\bibitem{contreras11}
P. C. Petersen \textit{et al}. Phys. Rev. Lett. {\bf 57}, 949 (1986).

\bibitem{milton}
M. D. Slaughter, Phys. Rev. C {\bf 82}, 015208 (2010).

\bibitem{milton1}
C. Amsler \textit{et al}. Phys. Lett. B {\bf 667}, 1 (2008); also 2009 partial update for the 2010 edition.

\bibitem{contreras12}
H. T. Diehl  \textit{et al}. Phys. Rev. Lett. {\bf 67}, 804 (1991).

\bibitem{contreras13}
N. B. Wallace \textit{et al}. Phys. Rev. Lett. {\bf 74}, 3732 (1995).

\bibitem{contreras14}
A. Bosshard  \textit{et al}. Phys. Rev. D {\bf 44}, 1962 (1991).

\bibitem{aliev(2)}
T. M. Aliev, M. Savci, Phys. Rev. D {\bf 90}, 116006 (2014).

\bibitem{aliev22}
B. Krusche, S. Schadmand, Prog. Part. Nucl. Phys. {\bf 51}, 399 (2003).

\bibitem{aliev23}
M. Kotulla \textit{et al}. Phys. Rev. Lett. {\bf 89}, 272001 (2002).

\bibitem{aliev24}
M. Kotulla, Prog. Part. Nucl. Phys. {\bf 61}, 147 (2008).

\bibitem{aliev25}
V. Punjabi \textit{et al}. Phys. Rev. C {\bf 71}, 055202 (2005).

\bibitem{aman12}
X. Song, V. Gupta, Phys. Rev. D. {\bf 49}, 2211 (1994).



\bibitem{aman14}
M. D. Slaughter, Phys. Rev. C {\bf 82}, 015208 (2010).

\bibitem{aman15}
M. D. Slaughter, Phys. Rev. D {\bf 84}, 071303 (2011).

\bibitem{aman16}
J. Linde, T. Ohlsson, H. Snellman, Phys. Rev. D {\bf 57}, 452 (1998).




\bibitem{aman13}
H. Dahiya, M. Gupta, Phys. Rev. D {\bf 67}, 114015 (2003).

\bibitem{aman18}
I. S. Sogami, N. Oh'yamaguchi, Phys. Rev. Lett. {\bf 54}, 2295 (1985).





\bibitem{aman20}
F. Schlumpf, Phys. Rev. D {\bf 48}, 4478 (1993).

\bibitem{aman21}
G. Ramalho, K. Tsushima, F. Gross, Phys. Rev. D {\bf 80}, 033004 (2009).




 
\bibitem{aman27} 
F. X. Lee, Phys. Rev. D {\bf 57}, 1801 (1998).

\bibitem{aman28}
S. L. Zhu, W. Y. P. Hwang, Z. S. P. Yang, Phys. Rev. D {\bf 57}, 1527 (1998).

\bibitem{aman29}
A. Iqubal, M. Dey, J. Dey, Phys. Lett. B {\bf 477}, 125 (2000).

\bibitem{aman26}
T. M. Aliev, A. Ozpineci, M. Savci, Phys. Rev. D {\bf 62}, 053012 (2000).

\bibitem{aman30}
B. Schwesinger, H. Weigel, Nucl. Phys. A {\bf 540}, 461 (1992).

\bibitem{aman31}
Y. Oh, Phys. Rev. D {\bf 75}, 074002 (2007).

\bibitem{aman24}
B. S. Bains, R. C. Verma, Phys. Rev. D {\bf 66}, 114008 (2002).

\bibitem{aman25}
R. Dhir, R. C. Verma, Eur. Phys. J. A {\bf 42}, 243 (2009).

\bibitem{aman35}
R. Flores-Mendieta, Phys. Rev. D {\bf 80}, 094014 (2009).

\bibitem{aman36}
L. S. Geng, J. M. Camalich, M. J. V. Vacas, Phys. Rev. D. {\bf 80}, 034027 (2009).

\bibitem{aman32}
T. Ledwig, A. Silva, M. Vanderhaeghen, Phys. Rev. D {\bf 79}, 094025 (2009).

\bibitem{aman33}
 H-C. Kim, M. Praszalowicz, K. Goeke, Phys. Rev. D {\bf 57}, 2859 (1998).
 
\bibitem{aman34} 
G. S. Yang, H-C. Kim, M. Praszalowicz, K. Goeke, Phys. Rev. D {\bf 70}, 114002 (2004).



\bibitem{aman37}
S. Boinepalli, D. B. Leinweber, P. J. Moran, A. G. Williams, J. M. Zanotti, J. B. Zhang, Phys. Rev. D {\bf 80}, 054505 (2009).

\bibitem{aman38}
C. Aubin, K. Orginos, V. Pascalutsa, M. Vanderhaeghen, Phys. Rev. D {\bf 79}, 051502 (2009).

\bibitem{aman39}
P. E. Shanahan \textit{et al}. Phys. Rev. D {\bf 89}, 074511 (2014).

\bibitem{aman40}
F. X. Lee, R. Kelly, L. Zhou, W. Wilcox, Phys. Lett. B {\bf 627}, 71 (2005).

\bibitem{friese} 
V. Friese, Nucl. Phys. A {\bf 774}, 377 (2005).

\bibitem{gsi} 
$http://www. gsi. de/fair/index.html.$

\bibitem{zhan} 
W. Zhan \textit{et al.}, Int. J. Mod. Phys. E {\bf 15}, 1941 (2006).

\bibitem{yano} 
Y. Yano, Nucl. Instrum. Methods B {\bf 261}, 1009 (2007).

\bibitem{ganil} 
$http://www. ganil. spiral2.eu/research/developments/spiral2/.$

\bibitem{frib} 
Whitepapers of the 2007 NSAC Long Range Plan Town Meetings, January 2007 Chicago, $http://dnp. aps. org.$


\bibitem{azizi}
K. Azizi, G. Bozkir, Eur. Phys. J. C {\bf 76}, 521 (2016).

\bibitem{ryu1}
 C. Y. Ryu and K. S. Kim, Phys. Rev. C {\bf 82}, 025804 (2010).
 
  \bibitem{ryu2}
 C. Y. Ryu, M. K. Cheoun, C. H. Hyun, Journal of Korean Physical Society {\bf 54}, 1448 (2009).

 
\bibitem{happy}
H. Singh, A. Kumar, H. Dahiya, Chinese Physics C, {\bf 41}, 094104 (2017).

\bibitem{happy1}
H. Singh, A. Kumar, H. Dahiya, arxiv:1710.08328v1 [nucl-th].

\bibitem{serot2} 
R. J. Furnstahl, B. D. Serot, Phys. Rev. C {\bf 41}, 262 (1990).

 \bibitem{sw}
S. Weinberg, Physica A {\bf 96}, 327 (1979).

\bibitem{am}
A. Manohar, H. Georgi, Nucl. Phys. B  {\bf 234}, 189 (1984).

\bibitem{chengsu3}
  T. P. Cheng, L. F. Li, Phys. Rev. D {\bf 57}, 344 (1998).
  
\bibitem{cheng1} 
 T. P. Cheng, L. F. Li, Phys. Rev. Lett. {\bf 80}, 2789 (1998).
 
 \bibitem{song} 
H. Q. Song, R. K. Su, Phys. Lett. B {\bf 358}, 179 (1995).

\bibitem{cheng}  
T. P. Cheng, L. F. Li, Phys. Rev. Lett. {\bf 74}, 2872 (1995).

\bibitem{har}
H. Dahiya, M. Gupta, Phys. Rev. D {\bf 64}, 014013 (2001).

\bibitem{nmc}
New Muon Collaboration, P. Amaudruz \textit{et al.}, Phys. Rev. Lett. {\bf 66}, 2712 (1991).

\bibitem{aarti}
A. Gridhar, H. Dahiya, M. Randhawa, Phys.Rev. D {\bf 92} 3, 033012 (2015).


\bibitem{har2}
I. S. Sogami, N. Oh'yamaguchi, Phys. Rev. Lett {\bf 54}, 2295 (1985); Kuang-Ta Chao, Phys. Rev. D {\bf 41}, 920 (1990). 

\bibitem{gupta}
M. Gupta, J. Phys. G {\bf 16}, L 213 (1990).


\bibitem{wang}
P. Wang, Z. Zong-Ye and Y. You-Wen, Commun. Theor. Phys. {\bf 36}, 71 (2001).

\bibitem{wang2}
P. Wang \textit{et al.}, Phys. Rev. C {\bf 70}, 015202 (2004).

\bibitem{wang3}
P. Wang, Z. Y. Song \textit{et al.}, Nucl. Phys. A {\bf 688}, 791 (2001).
 
\bibitem{wang4}
P. Wang \textit{et al.}, Nucl. Phys. A {\bf 744}, 273 (2004). 

 \bibitem{barik1}
 N. Barik, B. K. Dash, Phys. Rev. D {\bf 31}, 7 (1985).
 
  \bibitem{barik2}
 N. Barik \textit{et al.}, Phys. Rev. C {\bf 88}, 015206 (2013).
 



 	
\bibitem{abu}
M. Abu-Shady, A. K. Abu-Nab, American Journal of Physics and App. {\bf 46}, 1 (2014). 

\bibitem{experiment} 
K. Hagiwara \textit{et al.}, Phys. Rev. D {\bf 66}, 010001 (2002).	

\bibitem{aarti18}
T. Ledwig, A. Silva, M. Vanderhaeghen, Phys. Rev D {\bf 79}, 094025 (2009).



\bibitem{aarti20}
R. Flores-Mendieta, Phys. Rev. D {\bf 80}, 094014 (2009); L. S. Geng, J. M. Camalich, M. J. V. Vacas, Phys. Rev D {\bf 80}, 034027 (2009).

\bibitem{aarti22}
F. X. Lee, R. Kelly, L. Zhou, W. Wilcox, Phys. Lett. B {\bf 627}, 71 (2005).

\bibitem{aarti23}
S. T. Hong, Phys. Rev. D {\bf 76}, 094029 (2007).


\bibitem{aarti24}
E. E. Jenkins, A. V. Manohar, Phys. Lett. B {\bf 335}, 452 (1994); M. A. Luty, J. March-Russell, M. J. White, Phys. Rev. D {\bf 51}, 2332 (1995); A. J. Buchmann, J. A. Hester, R. F. Lebed,
Phys. Rev. D {\bf 66}, 056002 (2002); A. J. Buchmann, R. F. Lebed, Phys. Rev. D {\bf 67}, 016002
(2003).

\bibitem{aarti26}
A. J. Buchmann, E. Hernandez, A. Faessler, Nucl. Phys. A {\bf 569}, 661 (1994); G. Wagner,
A. J. Buchmann, A. Faessler, Phys. Lett. B {\bf 359}, 288 (1995); A. J. Buchmann, E. Her-
nandez, A. Faessler, Phys. Rev. C {\bf 55}, 448 (1997); G. Wagner, A. J. Buchmann, A.
Faessler, Phys. Rev. C {\bf 58}, 3666 (1998); G. Wagner, A. J. Buchmann, A. Faessler, J. Phys.
G {\bf 26}, 267 (2000); A. J. Buchmann, E. M. Henley, Phys. Rev. C {\bf 63}, 015202 (2000); A. J.
Buchmann, Phys. Rev. Lett. {\bf 93}, 212301 (2004).

\bibitem{AMU}
A. Mishra \textit{et al.}, Eur. Phys, J. A {\bf 41}, 205  (2009).

 \bibitem{wang neu}
 P. Wang \textit{et al.}, Phys. Rev. C {\bf 72}, 045801 (2005).

\bibitem{mg} 
M. Gupta, N. Kaur, Phys. Rev. D {\bf 28}, 534 (1983).

\bibitem{thomas}
 S. Theberge, A. W. Thomas, Phys. Rev. D {\bf 25}, 284 (1982); 
J. Cohen, H. J. Weber, Phys. Lett. B {\bf 165}, 229 (1985).

\bibitem{rhic} 
$https://www. bnl. gov/rhic.$

\bibitem{nica} 
$https://nica. jinr. ru.$

\bibitem{base} 
$https://home. cern/about/experiments/base.$

		 
\end{thebibliography}
\end{document}